quant-ph/0501068

 \documentclass[aps,eqsecnum,preprint,showpacs]{revtex4}
 \usepackage{graphicx}

\begin{document}

 \def\BE{\begin{equation}}
 \def\EE{\end{equation}}
 \def\BA{\begin{array}}
 \def\EA{\end{array}}
 \def\BEA{\begin{eqnarray}}
 \def\EEA{\end{eqnarray}}
 \def\ds{\displaystyle}
 \def\r{\vec{\rho}}
 \def\q{\vec{q}}

 \title{Quantum parallel dense coding of optical images}

 \author{T. Yu. Golubeva$^{1}$, Yu. M. Golubev$^{1}$, I.~V.~Sokolov$^{1}$,
  and M.~I.~Kolobov$^{2}$}
 \affiliation{1 V.~A.~Fock Physics Institute, St. Petersburg University,
       198504 Petrodvorets, St.-Petersburg, Russia}
 \affiliation{2 Laboratoire PhLAM, Universit\'e de Lille-1,
       F-59655 Villeneuve d'Ascq cedex, France}

 \begin{abstract}

We propose quantum dense coding protocol for optical images. This protocol
extends the earlier proposed dense coding scheme for continuous variables
[S.~L.~Braunstein and H.~J.~Kimble, Phys.~Rev.~A {\bf 61}, 042302 (2000)]
to an essentially multimode in space and time optical quantum
communication channel. This new scheme allows, in particular, for parallel
dense coding of non-stationary optical images. Similar to some other
quantum dense coding protocols, our scheme exploits the possibility of
sending a classical message through only one of the two entangled
spatially-multimode beams, using the other one as a reference system. We
evaluate the Shannon mutual information for our protocol and find that it
is superior to the standard quantum limit. Finally, we show how to
optimize the performance of our scheme as a function of the
spatio-temporal parameters of the multimode entangled light and of the
input images.

 \end{abstract} 
 \maketitle


 \newpage


 \section{Introduction}

The fundamental properties and possible applications of non-classical light have
been extensively investigated starting from the mid-70s. In the past decade such
novel fields of application of non-classical light have emerged, as quantum
information \cite{Bouwmeester00,Kluwer03} and quantum imaging
\cite{Kolobov99,Lugiato02}. Quantum imaging uses spatially multimode
non-classical states of light with quantum fluctuations suppressed not only in
time, but also in space. The promising idea is to introduce optical parallelism
inherent in quantum imaging into various protocols of quantum information, such
as quantum teleportation, quantum dense coding, quantum cryptography etc., thus
increasing their information possibilities. The continuous variables quantum
teleportation protocol proposed in \cite{Braunstein98a, Bouwmeester97} and
experimentally realized in \cite{Furusawa98, Bowen03} has been recently extended
for teleportation of optical images in \cite{Sokolov01, Gatti04}.

Quantum dense coding has been firstly proposed and experimentally realized
for discrete variables, {\it qubits}, \cite{Bennett92, Mattle96} and later
generalized for continuous variables in \cite{Braunstein00}. In this paper
we propose the continuous variables quantum dense coding protocol for
optical images. Our scheme extends the protocol \cite{Braunstein00} to the
essentially multimode in space and time optical communication channel.
This generalization exploits the inherent parallelism of optical
communication and allows for simultaneous parallel dense coding of an
input image with $N$ elements. In the case of a single spatial mode
considered in \cite{Braunstein00} one has $N=1$.

We calculate the Shannon mutual information for a stream of classical
input images in coherent state. In this paper we assume arbitrarily large
transverse dimensions of propagating light beams and the unlimited spatial
resolution of photodetection scheme. That is, we actually find an upper
bound on the spatio-temporal {\it density} of the information stream in
bits per $cm^2 \cdot s$. This density depends on the degree of squeezing
and entanglement in non-classical illuminating light. Two sets of
spatio-temporal parameters play an important role in our protocol: i) the
coherence length and the coherence time of spatially-multimode squeezing
and entanglement, and ii) the spatio-temporal parameters of the stream of
input images. In our analysis we assume that the sender (Alice) produces a
uniform ensemble of images with Gaussian statistics, characterized by
certain resolution in space and time (the Alice's grain).

We demonstrate that the essentially multimode quantum communication channel
provides much higher channel capacity than a single-mode quantum channel due to
its intrinsic parallel nature. The density of the information stream is in
particular limited by diffraction. We find that the role of diffraction can be
partially compensated compensation by lenses properly inserted in the scheme. An
important difference between the classical communication channel (i.~e.~with
vacuum fluctuations at the input of the scheme instead of multimode entangled
light) and its quantum counterpart is that in quantum case there exists an
optimum spatial density of the signal image elements, which should be matched
with the spatial frequency band of entanglement.

In Sec.~II we describe in detail the scheme of the dense coding protocol for
optical images. The channel capacity of our communication scheme is evaluated in
Sec.~III. We make our conclusions in Sec.~IV. The relevant to our analysis
properties of spatially-multimode squeezing are given in Appendix.

 \section{Spatially-multimode quantum dense
coding channel}

The optical scheme implementing the protocol is shown in Fig.~1. Compared
to the generic continuous variables dense coding scheme
\cite{Braunstein00}, here the light fields are assumed to be
spatially-multimode.
 \begin{figure}[h]
 \includegraphics[width=140mm, angle=00]{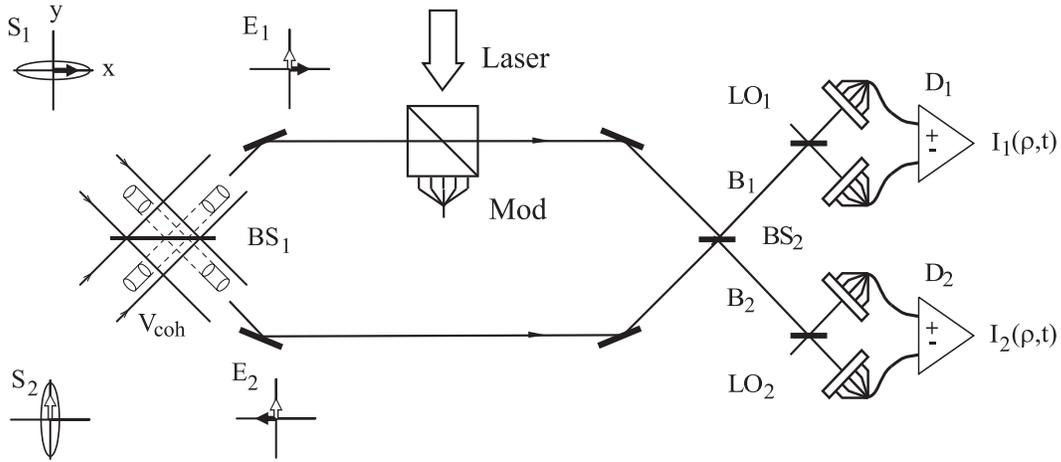}
 \label{schematic}
 \caption{Optical scheme for spatially-multimode dense coding.}
 \end{figure}
At the input, the spatially-multimode squeezed light beams with the slow
field amplitudes $S_1(\r,t)$ and $S_2(\r,t)$ in the Heisenberg
representation, are mixed at the symmetrical beamsplitter $BS_1$. For
properly chosen orientation of the squeezing ellipses of the input fields
the scattered fields $E_1(\r,t)$ and $E_2(\r,t)$ are in the entangled
state with correlated field quadrature components, as illustrated in
Fig.~1.

The classical signal image field $A(\r,t)$ is created by Alice in the
first beam by means, e.~g., of the controlled (with given resolution in
space-time) mixing device $Mod$ with almost perfect transmission for the
non-classical field $E_1(\r,t)$. The receiver (Bob) detects the entangled
state of two beams by means of optical mixing on the symmetrical output
beamsplitter $BS_2$ and the homodyne detection of quadrature components of
the output fields $B_1(\r,t)$ and $B_2(\r,t)$. This allows for measurement
of both quadrature components of the image field with effective quantum
noise reduction.

One can give a more straightforward explanation of the sub-shot-noise
detection of the signal in the scheme shown in Fig.~1. For the symmetrical
scattering matrix of the beamsplitters
 \BE
      \left\{ R_{nm} \right\} = \frac{1}{\sqrt{2}} \left(
      \begin{array}{rcr}
        1 && 1 \\
        1 && -1 \\
      \end{array}
      \right),
                     \label{beamsplitter}
 \EE
and equal optical paths of two beams, the effective Mach-Zehnder
interferometer directs the input squeezed field $S_1(\r,t)$ onto the
detector $D_1$, and similar for $S_2(\r,t)$, thus allowing for
sub-shot-noise detection of the squeezed quadrature components in both
beams.

The fields at the inputs of the homodyne detectors $D_1$ and $D_2$ are
 \BE
 B_n(\r,t) = S_n(\r,t) + \frac{1}{\sqrt{2}} A(\r,t),
 \label{output_field}
 \EE
where $n=1,2$. In the paraxial approximation, the slow amplitude of light
field $B_n(\r,t)$ is related to the creation and annihilation operators
$b_n^\dag(\q,\Omega)$ and $b_n(\q,\Omega)$ for the plane waves with the
transverse component of the wave vector $\q$ and frequency $\Omega$ by
 \BE
 B_n(\r,t)=\frac{1}{\sqrt{L^2 T}}\sum_{\q,\Omega} b_n(\q,\Omega)
 e^{i(\q\cdot\r-\Omega t)}.
 \label{Fourier_discrete}
 \EE
In the case of large quantization volume with the transverse and
longitudinal dimensions $L$ and $cT$, the summation is performed over the
following values of $\q$ and $\Omega$: $\q=(q_x,q_y)$, $\displaystyle{q_x
=\frac{2\pi}{L}n_x, q_y =\frac{2\pi}{L}n_y}$ and $\displaystyle{\Omega
=\frac{2\pi}{T}n}$ with $n_x, n_y$ and $n$ taking the values $0,\pm 1,\pm
2, \dots$.

The free-field commutation relations are given by
 \BEA
 \left[B_n(\r,t),B^\dag_{n'}(\r\;',t')\right] &=&
 \delta_{n,n'}\delta(\r-\r\;')\,\delta(t-t'), \nonumber \\
 \left[b_n(\q,\Omega),b_{n'}^\dag(\q\;',\Omega')\right] &=&
 \delta_{n,n'}\,\delta_{\q,\q\;'}\,\delta_{\Omega,\Omega'}.
 \label{commutators}
 \EEA
The value of the irradiance (in photons per $cm^2 \cdot s$) is equal to
$B_n^\dag(\r,t)B_n(\r,t)$, and the number of photons in the field mode
($\q,\Omega$), localized in the quantization volume $L^2 cT$, is
$b^\dag_n(\q,\Omega)b_n(\q,\Omega)$. The observed photocurrent densities
 \BEA
    I_1(\r,t) &=&  B_0 \big[B_1(\r,t) + B^{\dag}_1(\r,t)\big],
    \nonumber \\
    I_2(\r,t) &=&  B_0 \frac{1}{i} \big[B_2(\r,t) -
    B^{\dag}_2(\r,t)\big],
    \label{currents}
 \EEA
have the following Fourier amplitudes
 \BEA
 i_1(\q,\Omega) &=&  B_0 \Big[b_1(\q,\Omega) +  b_1^\dag(-\q,-\Omega) \Big],
       \nonumber \\
 i_2(\q,\Omega) &=&  B_0 \frac{1}{i}\Big[b_2(\q,\Omega) -
 b_2^\dag(-\q,-\Omega)\Big],
 \label{currents_Fourier}
 \EEA
where $B_0$ (taken as real) and $iB_0$ are the local oscillator amplitudes
used in the homodyne detection  (see the discussion in Subsection
\ref{subsection_Shannon}). Here and in what follows we denote the Fourier
amplitudes of the fields and the photocurrent densities by the lower-case
symbols.

The squeezing transformation performed by the optical parametric
amplifiers (OPAs), illuminating the inputs of the scheme, can be
written \cite{Kolobov99} as follows:
\BE
      s_n(\vec{q},\Omega) = U_n(\vec{q},\Omega) c_n(\vec{q},\Omega) +
      V_n(\vec{q},\Omega) c_n^{\dag}(-\vec{q},-\Omega),
                      \label{squeezing}
\EE
where the coefficients $U_n(\vec{q},\Omega)$ and $V_n(\vec{q},\Omega)$
depend on the pump-field amplitudes of the OPAs, their nonlinear
susceptibilities and the phase-matching conditions (see Appendix for
definitions of the squeezing parameters). The input fields
$c_n(\q,\Omega)$ of the OPAs are assumed to be in vacuum state.

After some calculation we obtain for the Fourier amplitudes of the
photocurrent densities:
 \BE
 i_n(\q,\Omega) = B_0\left\{f_n(\q,\Omega)+a_n(\q,\Omega)\right\},
 \label{currents_final}
 \EE
 where
 $$
 f_1(\q,\Omega)=\left[e^{r_1(\q,\Omega)}\cos\psi_1(\q,\Omega) +
 ie^{-r_1(\q,\Omega)}\sin\psi_1(\q,\Omega)\right]
 e^{-i\phi_1(\q,\Omega)}c_1(\q,\Omega)\,+\\
 $$
 \BE
 \big[h.c., (\q,\Omega)
 \rightarrow (-\q,-\Omega)\big],
 \label{noise_1}
 \EE
 and
 $$
 f_2(\q,\Omega)=\left[e^{-r_2(\q,\Omega)}\cos\psi_2(\q,\Omega) +
 ie^{r_2(\q,\Omega)}\sin\psi_2(\q,\Omega)\right]
 e^{-i\phi_2(\q,\Omega)}c_2(\q,\Omega)\,+\\
 $$
 \BE
 \big[h.c., (\q,\Omega)
 \rightarrow (-\q,-\Omega)\big],
 \label{noise_2}
 \EE
 represent the quantum fluctuations of the fields at both photodetectors, and
 \BE
 \BA{rlc}
 a_1(\q,\Omega)&=&{\ds \frac{1}{\sqrt{2}}}
 \Big[a(\q,\Omega)+a^*(-\q,-\Omega)\Big],\vspace{3mm}\\
 a_2(\q,\Omega)&=&{\ds \frac{1}{i\sqrt{2}}}\Big[a(\q,\Omega)
 -a^*(-\q,-\Omega)\Big],
 \EA
 \label{signals}
 \EE
 are the detected by Bob components of the Alice's signal image.
 Here $a(\q,\Omega)$ is the Fourier transform of classical field
 $A(\r,t)$, defined in analogy to (\ref{Fourier_discrete}).

\section{Channel capacity}

\subsection{Degrees of freedom of the noise and the image field}

In order to estimate the channel capacity one has to define the degrees of
freedom of the noise and the signal in our spatially-multimode scheme.

We shall assume that all elements of the scheme: OPAs non-linear crystals,
beamsplitters, modulator and CCD matrices of detectors, have large
transverse dimensions. The squeezed light fields are the stationary in
time and uniform in the cross-section of the beams random variables. That
is, all correlation functions of these fields are translationally
invariant in the ${\r,t}$ space. For the observed photocurrent densities
this implies that any pair of the Fourier amplitudes (\ref{noise_1}) and
(\ref{noise_2}) for given ($\q,\Omega$) and ($-\q,-\Omega$) result from
squeezing of the input fields $c(\q,\Omega)$ and $c(-\q,-\Omega)$ and
therefore is independent of any other pair.

On the other hand, since the observed photocurrent densities are real the
Fourier amplitudes $i_n(\q,\Omega)$ and $i_n^\dag(-\q,-\Omega)$ are not
independent, while
 \BE
 i_n(\q,\Omega) = i_n^\dag(-\q,-\Omega).
 \label{current_conjugate}
 \EE
For this reason we consider as independent random variables only the noise
terms in Fourier amplitudes $i_n(\q,\Omega)$ for $\Omega>0$. The real and
imaginary parts of the complex amplitudes $i_n(\q,\Omega)$ for $\Omega>0$
are related to the amplitudes of the real photocurrent noise harmonics
$\sim\cos(\q\cdot\r-\Omega t)$ and $\sim\sin(\q\cdot\r-\Omega t)$,
directly recovered by Bob from his measurements.

The Fourier amplitudes of the photocurrent densities
(\ref{currents_final}) satisfy the relation (\ref{current_conjugate}) and
therefore it is sufficient to take into account only $\Omega>0$. The
random signal sent by Alice is assumed to be stationary and uniform in the
cross-section of the beams. The amplitudes $a_n(\q,\Omega)$ for
$\Omega>0$, $n=1,2$, are taken as independent complex Gaussian variables
with variance $ \sigma^A(\q,\Omega)$ depending on $(\q,\Omega)$. Since the
transformation (\ref{signals}) is unitary, the Fourier classical
amplitudes $a(\q,\Omega)$ for any ($\q,\Omega$) are also statistically
independent, and the quantity
 \BE
 \sigma^A(\q,\Omega)=\langle |a(\q,\Omega)|^2\rangle,
 \label{signal_variance}
 \EE
is the mean energy of Alice's signal wave $(\q,\Omega)$ in the
quantization volume, where $\sigma^A(\q,\Omega) =
\sigma^A(-\q,-\Omega)$. Here the statistical averaging is
performed with the Gaussian complex weight function
 \BE
 {\cal P}^A_{\q,\Omega}(a(\q,\Omega))= \frac{1}{\pi
 \sigma^A(\q,\Omega)}
 \exp\left\{-\frac{|a(\q,\Omega)|^2}{\sigma^A(\q,\Omega)}\right\}.
 \label{Gaussian}
 \EE

In what follows we assume Gaussian spectral profile of width $q_A$
for the ensemble of input images in spatial frequency domain,
 \BE
 \sigma^A(\q,\Omega) = (2\pi)^3 \frac{P}{\pi(q_A/2)^2}
 \exp\left(- \frac{q_x^2+q_y^2}{(q_A/2)^2}\right)\Pi(\Omega)
 ,\,\,\,
 \Pi(\Omega)=\left\{
 \BA{ll}
 1/\Omega_A &|\Omega|\leq\Omega_A/2,\\
 0 & |\Omega|>\Omega_A/2,
 \EA\right.
 \label{spectrum}
 \EE
and, for the sake of simplicity, the narrow rectangular spectral
profile $\Pi(\Omega)$ of width $\Omega_A$ and height $1/\Omega_A$
in the temporal frequency domain. Since
 \BE
 \sum_{\q,\Omega} \sigma_A(\q,\Omega)= L^2TP,
 \label{photon_density}
 \EE
the total average density of photon flux in the image field per
$cm^2\cdot s$ is $P$. The variances of the observables
$i_n(\q,\Omega)$ are finally found in the form
 \BE
 \langle \frac{1}{2}\left\{i_n(\q,\Omega),
 i_n^\dag(\q,\Omega)\right\}_+\rangle  =
 B_0^2\left[\sigma_n^{BA}(\q,\Omega)+\sigma^A(\q,\Omega)\right],
 \label{variances}
 \EE
where $\{\phantom{a},\phantom{l}\}_+$ denotes the anticommutator.
The quantum noise variances in both detection channels are given
by
 \BE
 \sigma_n^{BA}(\q,\Omega)= \langle \frac{1}{2}\left\{f_n(\q,\Omega),
 f_n^\dag(\q,\Omega)\right\}_+\rangle,
 \label{noise_variances}
 \EE
 \BE
 \sigma_1^{BA}(\q,\Omega)=  e^{\ds 2r_1(\q,\Omega)}\cos^2\psi_1(\q,\Omega)
 + e^{\ds -2r_1(\q,\Omega)}\sin^2\psi_1(\q,\Omega),
 \label{noise_variance_1}
 \EE
 \BE
 \sigma_2^{BA}(\q,\Omega)=  e^{\ds -2r_2(\q,\Omega)}\cos^2\psi_2(\q,\Omega)
 + e^{\ds 2r_2(\q,\Omega)}\sin^2\psi_2(\q,\Omega).
 \label{noise_variance_2}
 \EE
Using these results we can evaluate the Shannon mutual information for our
dense coding scheme.

 \subsection{Shannon mutual information for the
 spatially-multimode dense coding channel}

 \label{subsection_Shannon}

It is well known that in the case of single-mode squeezed light field the
statistics of its quadrature amplitudes are Gaussian and can be
characterized, e.~g., by a Gaussian weight function in the Wigner
representation. In the homodyne detection of squeezed light, the
statistics of the photocounts are also Gaussian due to the linear relation
between the field amplitude and the photocurrent density. The discussion
of the homodyne detection in terms of the characteristic function can be
found in \cite{Braunstein90}. Some considerations for the homodyne
detection of spatially multimode fields are presented in \cite{Gatti04}.

In our quantum dense coding scheme the statistically independent degrees
of freedom of the noise and the signal are labeled by the frequencies
$(\q,\Omega)$ for $\Omega>0$. One can consider our quantum channel as a
collection of the statistically independent parallel Gaussian
communication channels in the Fourier domain. The mutual information
between Alice and Bob for given detector and frequencies $(\q,\Omega)$ is
defined as
 \BE
 I^S_n(\q,\Omega)= H_n^B(\q,\Omega)-\overline{H_n^{(B|A)}(\q,\Omega)}^A.
 \label{information_def}
 \EE
Here $H^B(\q,\Omega)$ is the entropy of Bob's observable, and
 $$
 \overline{H_n^{(B|A)}(\q,\Omega)}^A
 $$
is the averaged over the ensemble of Alice's signals entropy of noise,
introduced by the channel \cite{Holevo82}. For a Gaussian channel the
mutual information is given by
 \BE
 I^S_n(\q,\Omega)=
 \ln\left(1+\frac{\sigma^A(\q,\Omega)}{\sigma_n^{BA}(\q,\Omega}\right).
 \label{information_Gaussian}
 \EE
The quantum noise suppression within the frequency range of
effective squeezing and entanglement increases the signal-to-noise
ratio at the right side of (\ref{information_Gaussian}). The total
mutual information $I^S$, associated with the large area $L^2$ and
the large observation time $T$, is defined as a sum over all
degrees of freedom and is related to the {\sl density of the
information stream $J$ in bits per $cm^2\cdot s$}:
 \BE
 I^S=\sum_{n,\q,\Omega>0} I^S_n(\q,\Omega) = L^2T\,J,
 \label{information_result}
 \EE
where
 \BE
 J = \frac{1}{(2\pi)^3}\int d\q \int_{\Omega>0} d\Omega
 \sum_{n=1,2} I^S_n(\q,\Omega).
 \label{information_density}
 \EE

For qualitative and numerical analysis it is natural to associate such
quantities as the density of the information stream and of the photon flux
with the physical parameters present in our quantum dense coding scheme.
Squeezing and entanglement produced by the type-I optical parametric
amplifiers (OPA's), are characterized by the effective spectral widths
$q_c$ and $\Omega_c$ in the spatial and temporal frequency domain. The
coherence area in the cross-section of the beams and the coherence time
are introduced as $S_c = (2\pi/q_c)^2$ and $T_c= 2\pi/\Omega_c$. For
simplicity we assume that both OPA's have the same coherence area and
coherence time. The correlation area $S_A$ and the correlation time $T_A$
of non-stationary images sent by Alice, are related to the spectral widths
of the signal $q_A$ and $\Omega_A$ by $S_A = (2\pi/q_A)^2$ and $T_A=
2\pi/\Omega_A$. We consider the broadband degenerate collinear phase
matching in the traveling-wave type-I OPA's. The coherence time $T_c$ of
the spontaneous downconversion will be typically short compared to the
time duration $T_A$ of the Alice's movie frame.

The dimensionless information stream $\cal J$ and the
dimensionless input photon flux $\cal P$ are defined by $ {\cal J}
= S_c T_A J$, ${\cal P} = S_c T_A P$. That is, we relate both
quantities to the time duration of the Alice's movie frame and the
coherence area of squeezing and entanglement.

The optimum entanglement conditions in the OPA's are given by
 \BEA
      r_1(\vec{q},\Omega) &=& r_2(\vec{q},\Omega) \equiv r(\vec{q},\Omega),
                           \nonumber \\
       \psi_1(\vec{q},\Omega) &=& \psi_2(\vec{q},\Omega) \pm \pi/2 \equiv
       \psi(\vec{q},\Omega),\\
       \psi(0,0)&=&\pi/2. \nonumber
                      \label{orthogonal_ellipses}
 \EEA
We find the dimensionless information stream $\cal J$ in the following
form:
 \BE
 {\cal J} = \int d{\vec\kappa} \ln\left\{1+{\cal P}\frac{1}{
 \sigma^{BA}({\vec \kappa},0)}
 \frac{1}{\pi (d_A/2)^2}
 \exp\left(-\frac{\kappa_x^2+\kappa_y^2}{(d_A/2)^2}\right)\right\},
 \label{information_final}
 \EE
where
 \BE
\sigma^{BA}({\vec \kappa},0) = e^{\ds
2r({\vec\kappa},0)}\cos^2\psi({\vec\kappa},0) + e^{\ds
-2r({\vec\kappa},0)}\sin^2\psi({\vec\kappa},0),
 \EE
and dimensionless spatial frequency is defined as ${\vec \kappa} =
\q/q_c$. The relative spectral width of the Alice's signal $d_A = q_A/q_c
= (S_c/S_A)^{1/2}$ can be interpreted as the number of image elements per
coherence length, i.~e.~the relative linear density of image elements. In
what follows we assume a simple estimate $q_c/2 = \sqrt{2k/l}$, related to
the diffraction spread of parametric downconversion light inside the OPA
crystal, where $k$ is the wavenumber  and $l$ is the crystal length.

Quantum noise in the dense coding scheme is effectively reduced for
optimum phase matching of squeezed beams. As shown in
\cite{Kolobov89a,Kolobov89b,Kolobov99}, an important factor is the
spatial-frequency dispersion of squeezing, that is, the $\q$-dependence of
the phase of the squeezed quadrature component. This dependence is due to
the diffraction inside the OPA. A thin lens properly inserted into the
light beam can effectively correct the $\q$-dependent orientation of
squeezing ellipses, as illustrated in Fig.~2.
 \begin{figure}[h]
 \includegraphics[width=80mm, angle=00]{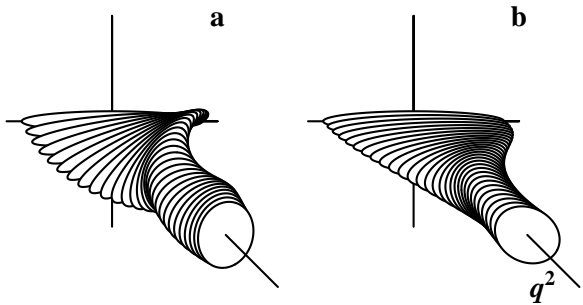}
 \label{squeezing_ellipses}
\caption{Spatial-frequency dependence of squeezing ellipses without (a)
and with (b) phase correction by means of a thin lens. The degree of
squeezing is $\exp[r(0,0)]=3$.}
 \end{figure}

The improvement in the signal-to-noise ratio for different spatial
frequencies can be characterized by the inverse noise variance
$\sigma^{BA}({\vec \kappa},0)$ shown in Fig.~3. As seen from this figure,
the phase correction by means of a lens allows for the low-noise signal
transmission within the spatial-frequency band of the effective squeezing.
 \begin{figure}[h]
 \includegraphics[width=80mm, angle=00]{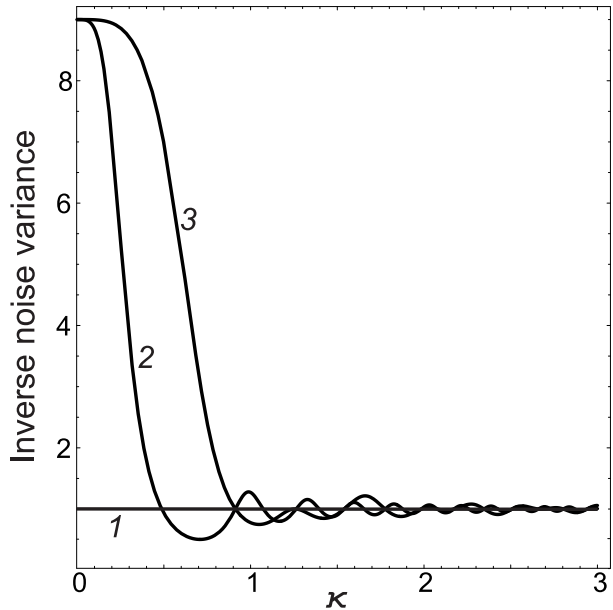}
 \label{inverse_variance}
 \caption{Inverse noise variance in dependence of the spatial
frequency $\kappa$ for vacuum noise at the input (1), and for squeezing
with $\exp[r(0,0)]=3$ without (2) and with (3) phase correction.}
 \end{figure}

In our plots for the mutual information density ${\cal J}$ we keep
constant the coherence area $S_c$, the degree of squeezing $r(0,0)$, and
the density of signal photons flux. The dependence of mutual information
density on the relative linear density $d_A$ of the image elements is
shown in Fig.~4.
 \begin{figure}[h]
 \includegraphics[width=80mm, angle=00]{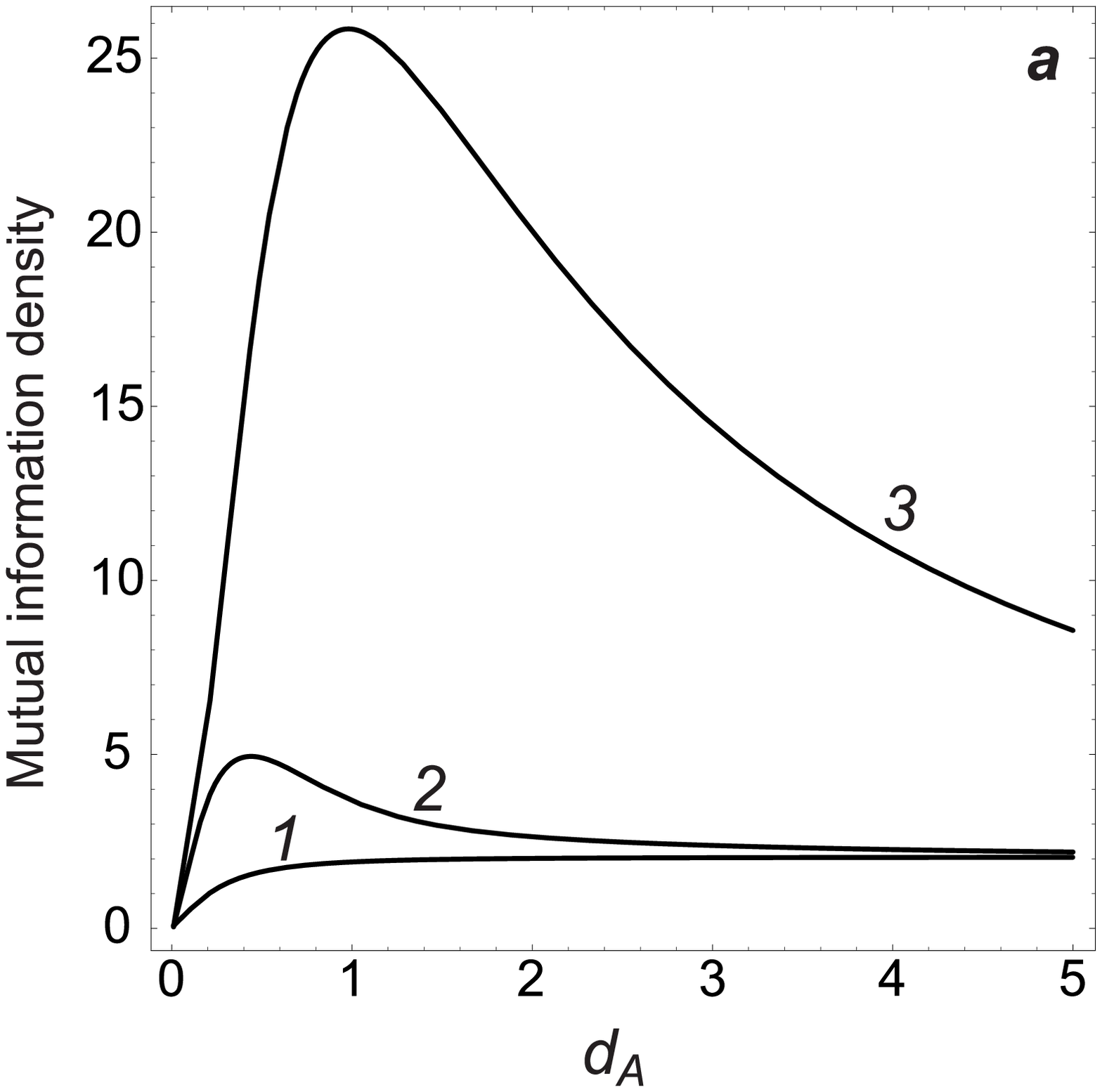}
 \includegraphics[width=80mm, angle=00]{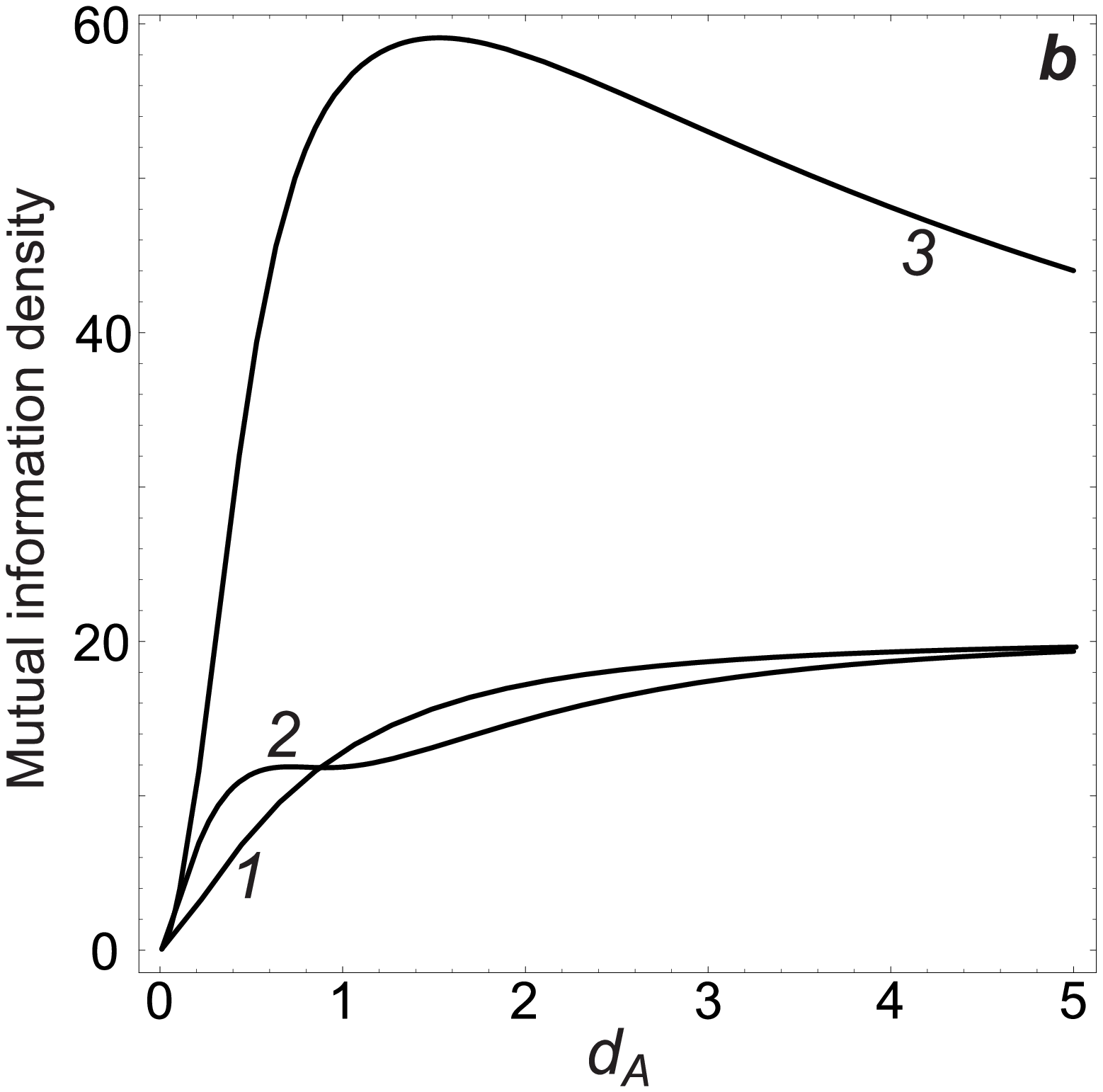}
 \label{mutual_information}
 \caption{Mutual information density for the vacuum noise at the input
of the scheme (1), and for squeezing with $\exp(r(0,0))=10$ without (2)
and with (3) phase correction. The density of signal photons is ${\cal
P}=1$ (Fig.4a), ${\cal P}=10$ (Fig.~4b).}
 \end{figure}
For $d_A \ll 1$ (large image elements, $S_A \gg S_c$), the mutual information
density increases linearly with $d_A$, since this means improvement of spatial
resolution in the input signal. Multimode quantum entanglement between two
channels of the scheme results in much higher channel capacity compared to the
classical limit (vacuum noise at the input of the scheme).

On the other hand, for $d_A \gg 1$ (image elements much smaller than the
coherence length), the effect of entanglement on the channel capacity is washed
out and ${\cal J}$ goes down to the vacuum limit. This is due to the fact that
in the limit $S_A \ll S_c$ almost all spatial frequencies of the signal are
outside the spatial-frequency band of the effective noise suppression, and the
channel capacity is finally limited by vacuum noise.

The phase correction of squeezing and entanglement significantly improves the
channel capacity, since it brings the spatial frequency band of the effective
noise suppression to its optimum value. It eliminates the destructive effect of
the amplified (stretched) quadrature of the noise field at the higher spatial
frequencies, as seen e.~g.~from Fig.~4b.

\section{Conclusion}

In this paper we have extended the continuous variables dense coding protocol
proposed in~\cite{Braunstein00} onto the optical images and calculated the {\sl
spatio-temporal density} of the Shannon mutual information. Our multimode
quantum communication channel provides much higher channel capacity due to its
intrinsic parallel nature. We have considered the role of diffraction in our
protocol and have found how to optimize its performance by means of a lens
properly inserted into the scheme. We have shown that, by contrast to the
classical communication channel, there exists an optimum spatial density of
image elements, matched to the spatial-frequency band of squeezing and
entanglement.

\section*{Acknowledgments}

The authors thank L.~A.~Lugiato and C.~Fabre for valuable discussions. This work
was supported by the Network QUANTIM (IST-2000-26019) of the European Union, by
the INTAS under Project 2001-2097, and by the Russian Foundation for Basic
Research under Project 03-02-16035. The research was performed within the
framework of GDRE "Lasers et techniques optiques de l'information".

\appendix
\section{Properties of spatially-multimode squeezing}

The main results for spatially-multimode squeezing are summarized in
\cite{Kolobov99}. The coefficients of the squeezing transformation
(\ref{squeezing}) satisfy the conditions
 \BEA
 & & |U_n(\q,\Omega)|^2 - |V_n(\q,\Omega)|^2 = 1,
                \label{conditions} \\
 & & U_n(\q,\Omega)V_n(-\q,-\Omega) = U_n(-\q,-\Omega)V_n(\q,\Omega),
                \nonumber
 \EEA
which are necessary and sufficient for preservation of the free-field
commutation relations (\ref{commutators}). The spatial and temporal
parameters of squeezed and entangled light fields essentially depend on
the orientation angle $\psi_n(\vec{q},\Omega)$ of the major axes of the
squeezing ellipses,
\BE
     \psi_n(\vec{q},\Omega) = \frac{1}{2}\arg\left\{U_n(\vec{q},\Omega)
     V_n(-\vec{q},-\Omega)\right\},
                       \label{psi}
\EE
and on the degree of squeezing $r_n(\vec{q},\Omega)$,
\BE
       e^{\pm r_n(\vec{q},\Omega)} = |U_n(\vec{q},\Omega)| \pm
       |V_n(\vec{q},\Omega)|.
                         \label{exponential}
\EE
The phase of the amplified  quadrature components is given by
\BE
     \phi_n(\vec{q},\Omega) = - \frac{1}{2}\arg\left\{U_n(\vec{q},\Omega)
     V_n^*(-\vec{q},-\Omega)\right\}.
                       \label{phi}
\EE
In analogy to the single-mode EPR beams, the multimode EPR beams are created if
squeezing in both channels is effective, and the squeezing ellipses are oriented
in the orthogonal directions.

For the type-I phase-matched traveling-wave OPAs, the coefficients
$U(\q,\Omega)$ and $V(\q,\Omega)$  are given by
 $$
        U(\q,\Omega) = \exp \Big\{i\Big[(k_z(\q,\Omega) - k)l -
        \delta(\q,\Omega)/2\Big]\Big\} \left[\cosh \Gamma(\q,\Omega) +
        \frac{i \delta(\q,\Omega)}{2
        \Gamma(\q,\Omega)} \sinh \Gamma(\q,\Omega)\right],
 $$
 \BE
        V(\q,\Omega) = \exp \Big\{i\Big[(k_z(\q,\Omega) - k)l -
        \delta(\q,\Omega)/2\Big]\Big\} \frac{g}{\Gamma(\q,\Omega)} \sinh
        \Gamma(\q,\Omega).
            \label{UandV}
 \EE
Here $l$ is the length of the nonlinear crystal, $k_z(\q,\Omega)$
is the longitudinal component of the wave vector ${\vec
k}(\q,\Omega)$ for the wave with frequency $\omega + \Omega$ and
transverse component $\q$. The dimensionless mismatch function
$\delta(\q,\Omega)$ is given by
 \BE
        \delta(\q,\Omega) = \Big(k_z(\q,\Omega)+k_z(-\q,-\Omega)-k_p\Big)l
        \approx (2 k-k_p)l+k''_\Omega l \Omega^2 - q^2 l/k,
             \label{mismatch}
 \EE
where $k_p$ is the wave number of the pump wave, $k_p-2 k = 0$ in the
degenerate case. We have assumed the paraxial approximation. The parameter
$\Gamma({\vec q},\Omega)$ is defined as
 \BE
         \Gamma(\q,\Omega) = \sqrt{g^2 -\delta^2(\q,\Omega)/4},
            \label{Gamma}
 \EE
where $g$ is the dimensionless coupling strength of nonlinear
interaction, taken as real for simplicity. It is proportional to
the nonlinear susceptibility, the length of the crystal, and the
amplitude of the pump field.

\bibliographystyle{plain}

\end{document}